\documentclass[a4paper,10pt]{article}
\usepackage{graphicx}
\usepackage{fancyheadings}
\begin{document}
\noindent
{\it 10th Hel.A.S Conference}\\
\noindent
{\it Ioannina, 5-8 September, 2011}\\
\noindent
%
%
CONTRIBUTED POSTER\\
\noindent
\underline{~~~~~~~~~~~~~~~~~~~~~~~~~~~~~~~~
~~~~~~~~}
\vskip 1cm
%
%
\begin{center}
{\Large\bf
The contact system DF~Hya revisited
}
\vskip 0.5cm
%
%
{\it
I. Kamenidis, A. Liakos and P. Niarchos
}\\
%
%
Department of Astrophysics, Astronomy and Mechanics, University of Athens, Athens, Hellas
\end{center}
\vskip 0.7cm
%
%
\noindent
{\bf Abstract: }
New BVRI CCD photometric observations of the contact system DF~Hya have been obtained. The light curves were analyzed with the Wilson-Devinney code and new geometric and photometric elements were derived. Moreover, the light curve solution, with the assumption of a third light in the system, revealed the existence of a tertiary component around the eclipsing pair. The present results are compared with those of other recent studies.


\section{Introduction, observations and analyses}
DF Hya is a contact system of W UMa-type with a period of 0.3306022$^d$. Various papers have been published for this system (cf. \cite{LI90};\cite{NI92};\cite{ZA09};\cite{XI09}) concerning both light curve and orbital period changes analyses.

The system was observed during the night of January 17 2011 at the Gerostathopoulion Observatory in Athens University using a 40-cm Cassegrain telescope equipped with the CCD camera ST-10XME and B, V, R, I Bessell photometric filters. Differential magnitudes were obtained with the software MUNIWIN v.1.1.26 \cite{HR98}, while GSC 0225-0943 and GSC 0225-0731 were used as comparison and check stars, respectively.

\begin{figure}[h]
\begin{tabular}{cc}
\centering
\includegraphics[width=7.5cm]{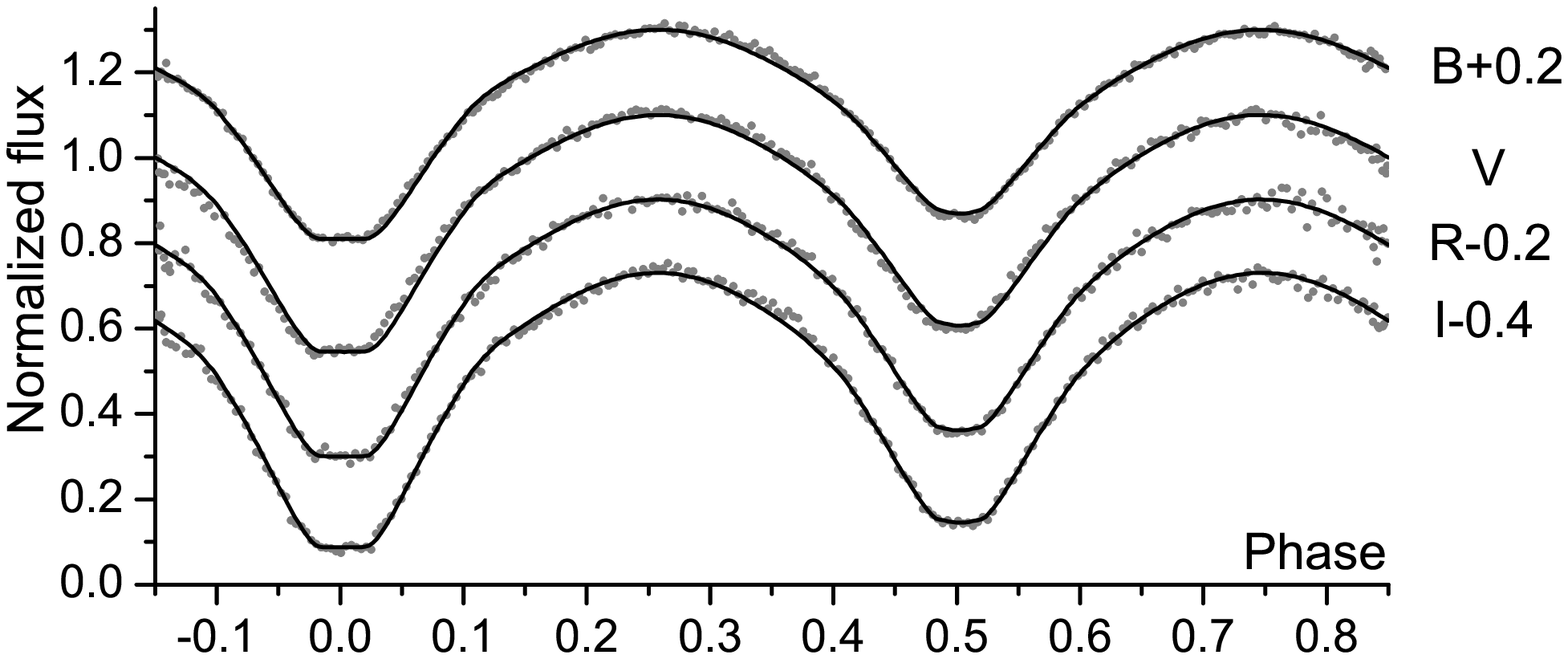}&\includegraphics[width=7.5cm]{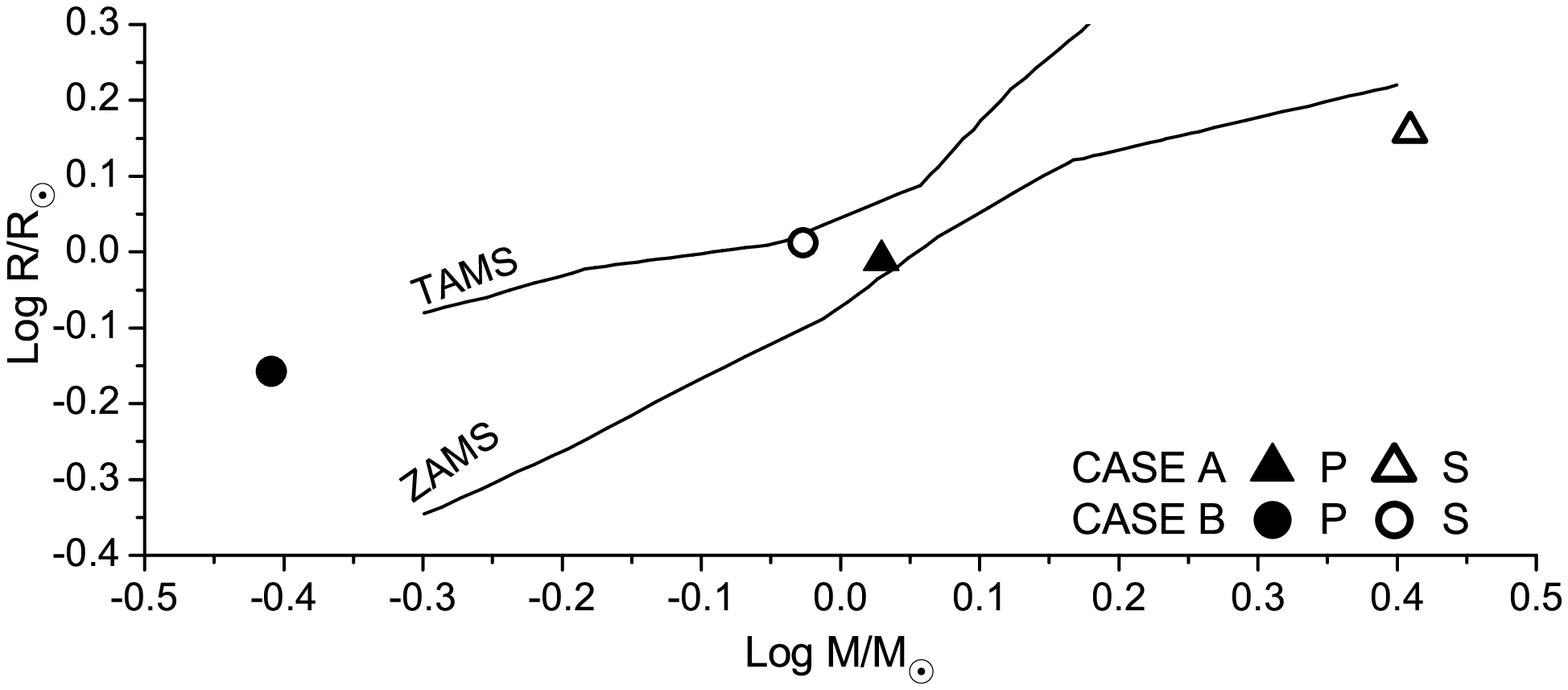}
\end{tabular}
\caption{Left panel: Observed (points) and synthetic (solid lines) light curves of DF~Hya. Right panel: The position of the system's components for Cases A and B. P and S refer to primary and secondary components, respectively. }
\label{fig1}
\end{figure}

The light curves were analysed with the PHOEBE v.0.29d software \cite{PZ05}. Mode 3 (contact system) was chosen for fitting applications and the `q search method' was applied for an estimation of the mass ratio with a step of 0.1 in the range of 0.1-10. This value then was adjusted in the subsequent analysis. The temperature of the primary component was set as a fixed parameter (T1=6000 K; \cite{MA06}), while the temperature of the secondary was left free. The albedos A$_1$, A$_2$  and gravity darkening coefficients g$_1$, g$_2$ were given theoretical values according to the components' spectral types. The potentials $\Omega_1$, $\Omega_2$, the system's inclination \textit{i} and the fractional luminosity of the primary component $L_1$ were also adjusted. The limb darkening coefficients $x_1,~x_2$  were taken from the tables of \cite{VH93}. Given the evidence for third body existence in the system (\cite{ZA09}; \cite{XI09}) the third light parameter $l_3$ was trialed. Synthetic and observed light curves are shown in Fig. \ref{fig1} with corresponding parameters given in Table \ref{tab1}.

The absolute parameters of the components were calculated (Table \ref{tab1}) and used for further study of their present evolutionary status. Two cases are considered: (\textbf{A}) The mass of the primary (hotter) and (\textbf{B}) the mass of the secondary (cooler) assigned values according to their spectral types as Main Sequence stars. The location of the components in a theoretical Mass-Radius diagram is illustrated in Fig. \ref{fig1} for both cases.

\begin{table}[h]

\caption{Results of light curve analysis and absolute parameters of the components (comp.).}
\label{tab1}
\scalebox{0.7}{

\begin{tabular}{lcc lcc cc l cc cc}
\hline
                        \multicolumn{8}{c}{\bf Light curve parameters}                        &      \multicolumn{5}{c}{\bf Absolute parameters}           \\
\hline
                &         &        &                &         &         &        &            &    & \multicolumn{2}{c}{Case A}& \multicolumn{2}{c}{Case B}\\
\cline{10-13}
\emph{Comp.:}   & \emph{P}&\emph{S}&\emph{Filters:} &\emph{B} & \emph{V}&\emph{R}& \emph{I}   &\emph{Comp.:}      & \emph{P}&\emph{S}& \emph{P}&\emph{S}   \\
\hline
T [K]           &   6000*  &5620 (5)&   $x_{1}$      &  0.709  &  0.572  &   0.490  &  0.411  &M [M$_{\odot}$]    & 1.07*   & 2.5 (2)& 0.39 (3)&  0.94*    \\
g               &    0.32* & 0.32*  &   $x_{2}$      &  0.764  &  0.623  &   0.537  &  0.452  &R [R$_{\odot}$]    & 0.97 (2)&1.44 (2)& 0.70 (2)& 1.03 (2)  \\
A               &    0.5*  & 0.5*   & L$_1$/L$_T$    &0.445 (5)&0.373 (4)&0.365 (2) &0.357 (1)&L [L$_{\odot}$]    & 1.10 (5)&2.41 (6)&0.56 (3) & 1.23 (3)  \\
$\Omega$        & 5.68 (1) &5.68 (1)& L$_2$/L$_T$    &0.509 (2)&0.604 (2)&0.617 (1) &0.628 (1)&M$_{bol}$ [mag]    &4.64 (2) &4.08 (2)&5.37 (2) & 4.81 (2)  \\
i [deg]&\multicolumn{2}{c}{84.8 (2)}& L$_3$/L$_T$    &0.046 (8)&0.023 (7)&0.17 (3)  &0.015 (3)&a [R$_{\odot}$]    & 2.17 (6)& 0.91(3)&1.56 (5) & 0.66 (2)  \\
q      &\multicolumn{2}{c}{2.38 (1)}&                &         &         &          &         &log$g$ [cm/s$^2$]  &4.49 (2)&4.53 (5) &4.34 (4) & 4.39 (1)  \\
\hline
 \multicolumn{13}{l}{*assumed, L$_T$=L$_1$ + L$_2$ +L$_3$, P=Primary, S= Secondary}
\end{tabular}}

\end{table}
\section{Discussion and conclusions}

New photometric study of the eclipsing binary DF~Hya was performed and new geometric elements of the system and absolute parameters of its components were derived. The system is of W-type, meaning that the primary component (hotter) is less massive and smaller than the secondary. The components' location in the M-R diagram (see Fig. \ref{fig1}) for Case A indicate that the secondary is a pre-Main Sequence star with a mass of 2.5 $M_\odot$, which is quite unusual for members of W~UMa systems. The second hypothesis (Case B) seems to be more likely as both stars lie inside the ZAMS-TAMS limits and therefore it is suggested as the most realistic one.
The comparison between the present light curves and the ones given by \cite{NI92} shows that brightness changes occur in the system affecting both the minima and maxima. In order to check any brightness variation periodicities, long term monitoring of the system is required.
For an accurate solution a third light was considered in the light curve analysis and a contribution of $\sim$2.5\% was found. According to the O-C analysis results of \cite{ZA09} (M$_{3,min}$=0.84 $M_\odot$), and assuming the MS nature of the tertiary component and taking into account the Mass-Luminosity relation for MS stars (L$\sim$M$^{3.5}$), the expected light contribution was found $\sim$25\%, which is much larger than the observed one. A possible explanation for this mismatch could be a binary star orbiting DF~Hya instead of a single one. On the other hand, the orbital period analysis of \cite{XI09} suggested a minimal mass of 0.21 $M_\odot$ for the additional component and mass transfer between the binary's members. This value does not satisfy also the observed light contribution, but if we assume a non coplanar orbit of the third body with an inclination of $\sim30^\circ$, then the result can be adopted. The Applegate mechanism was also tested for \cite{ZA09} results, but it was found that the quadrupole moment variation for both stars ($\Delta Q < 10^{50}~gr~cm^2$) cannot implicate the period changes \cite{LR02}. Hence, the third body existence remains as the most possible solution for the orbital period modulations of the system.
Spectroscopic observations are needed in order to check the value of the present photometric mass ratio and provide the information for a more accurate determination of the absolute parameters.

\end{document}